\documentclass[12pt, a4paper]{revtex4-2} % for arXiv
\usepackage{graphicx}% Include figure files
\usepackage{dcolumn}% Align table columns on decimal point
\usepackage{bm}% bold math
\usepackage{xspace}% To avoid problems with missing or double spaces after
\usepackage{lineno}% for linenumbers
\usepackage{multirow}

\usepackage{hyperref}% add hypertext capabilities
\def\sqsnn      {{\ensuremath{\protect\sqrt{s_{NN}}}}\xspace}
\def\gev        {{\ensuremath{\mathrm{\,Ge\kern -0.1em V}}}\xspace}
\def\gevc       {{\ensuremath{{\mathrm{\,Ge\kern -0.1em V\!/}c}}}\xspace}
\def\gevcc      {{\ensuremath{{\mathrm{\,Ge\kern -0.1em V\!/}c^2}}}\xspace}
\def\tev        {{\ensuremath{\mathrm{\,Te\kern -0.1em V}}}\xspace}
\def\cquark     {{\ensuremath{c}}\xspace}
\def\cquarkbar  {{\ensuremath{\overline \cquark}}\xspace}
\def\ccbar      {{\ensuremath{\cquark\cquarkbar}}\xspace}

\def\PJ         {\ensuremath{J}\xspace}                 
\def\Ppsi       {\ensuremath{\psi}\xspace}                 
\def\jpsi       {{\ensuremath{{\PJ\mskip -3mu/\mskip -2mu\Ppsi\mskip 2mu}}}\xspace}

\def\psip       {{\ensuremath{\Ppsi{'}}}\xspace}

\begin{document}
%\preprint{APS/123-QED}

\title{Towards quarkonium formation time determination}% Force line breaks with \\

\author{E. G. Ferreiro$^{a}$, F. Fleuret$^b$, E. Maurice$^b$}
\affiliation{%
$^a$Departamento de F\'isica de Part\'iculas and IGFAE, Universidade de Santiago de Compostela, 15782 Santiago de Compostela, Spain\\
$^b$Laboratoire Leprince-Ringuet, CNRS/IN2P3, \'Ecole polytechnique, Institut Polytechnique de Paris, Palaiseau, France\\
%\\
}%
\date{\today}% It is always \today, today,
\begin{abstract}
We propose a parametrization of the nuclear absorption mechanism relying on the proper time spent by \ccbar bound states travelling in nuclear matter. Our approach could lead to the extraction of charmonium formation time. It is based on a large amount of proton-nucleus data, from nucleon-nucleon center-of-mass energies $\sqsnn=27\gev$ to $\sqsnn=5.02\tev$, collected in the past 30~years, and for which the main effect on charmonium production must be its absorption by the nuclear matter it crosses.
\end{abstract}
\maketitle
%\linenumbers

%\title{Towards quarkonium formation time determination}% Force line breaks with \\
%\author{E. G. Ferreiro\inst{1} \and F. Fleuret\inst{2} \and E. Maurice\inst{2}}
%\institute{%
%Departamento de F\'isica de Part\'iculas and IGFAE, Universidade de Santiago de Compostela, 15782 Santiago de Compostela, Spain \and
%Laboratoire Leprince-Ringuet, CNRS/IN2P3, \'Ecole polytechnique, Institut Polytechnique de Paris, Palaiseau, France\\
%}%
%\abstract{
%We propose a parametrization of the nuclear absorption mechanism relying on the proper time spent by \ccbar bound states travelling in nuclear matter. Our approach could lead to the extraction of charmonium %formation time. It is based on a large amount of proton-nucleus data, from nucleon-nucleon center-of-mass energies $\sqsnn=27\gev$ to $\sqsnn=5.02\tev$, collected in the past 30~years, and for which the main %effect on charmonium production must be its absorption by the nuclear matter it crosses.}
%\maketitle

%\linenumbers

\begin{figure}[t]
\includegraphics[width=0.75\textwidth]{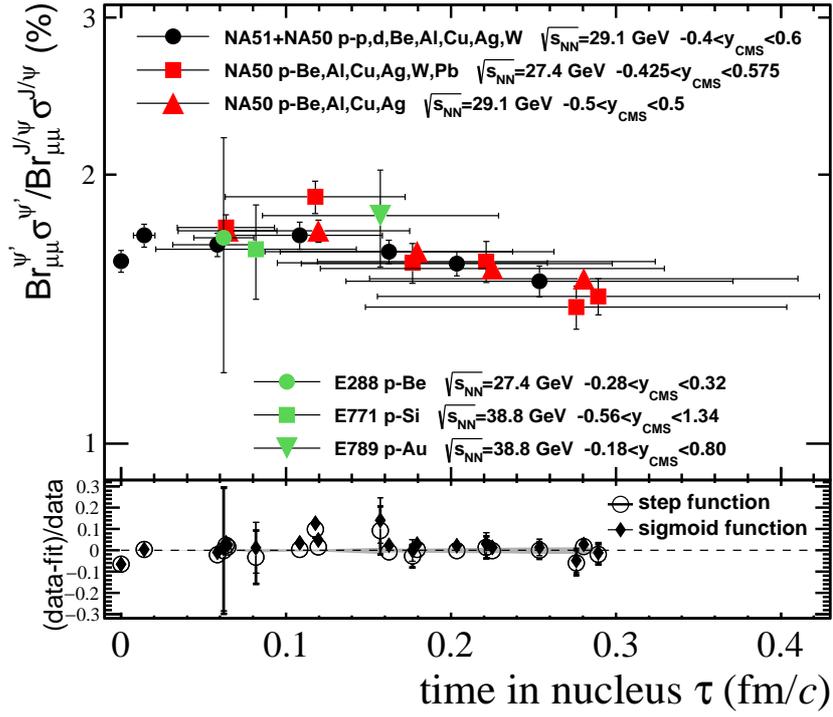}
\caption{\label{fig:Ratios} Top: \psip over \jpsi production ratios from \cite{NA51PLB438,NA50PLB553,NA50EPJC33,NA50EPJC48,E288PRL36,E771PLB374,E789PRD52} as a function of $\tau$, for various targets at various energies; Bottom: Relative difference between data and eq.~\ref{eq:abs2} step function (open circles), between data and eq.~\ref{eq:exact} sigmoid function (close diamonds), all parameters being determined with fits based on eq.~\ref{eq:abs2}. The Grey band shows the effects of $\tau_0$ uncertainties on the sigmoid function.}
\end{figure}

The production of charmonia, \cquark\cquarkbar bound states, is the object of forceful researches in proton-proton, proton-nucleus and nucleus-nucleus collisions. Their production is intrinsically a two-scale problem, that of the heavy-quark pair production, manageable with perturbative methods, and that of its hadronization, non-perturbative and due to the color confinement QCD property. Today, nearly all the models of charmonium production rely on a factorisation between the heavy-quark pair production and its hadronisation, the evolving heavy-quark pair being in a Color-Singlet (CS) or a Color-Octet (CO) state~\cite{JPLPR889}. In proton-nucleus collisions, several initial and final state effects, also called Cold Nuclear Matter (CNM) effects, can  modify charmonium yields. Charmonia can be suppressed due to nuclear absorption \cite{NAPLB207}, suffer multiple scatterings or lose their energy by radiation \cite{ELJHEP03}, in their way out of the nucleus overlapping region. They can also be broken by comovers \cite{CapellaPLB393}\cite{GavinPRL78}\cite{ArmestoPLB430}\cite{CapellaPRL85}  or be affected by the modification of the parton flux inside nuclei as encoded in nuclear PDFs \cite{PDFEPJC77}\cite{PDFPRD93}.
The relative importance of the above-cited effects depends essentially on the collision energy, the transverse momentum and the rapidity of the probe, together with the nuclear size \cite{NPB345,PRC87,LOUR09}. 
\\
In this Letter, we propose to exploit the charmonium nuclear absorption (or break-up) effect to explore the \ccbar hadronization mechanism. After its production, the small radius \ccbar pair is expected to bind into a larger radius colour neutral state \cite{KOPPRD44}\cite{BRODPLB206}. The latter may interact with the nucleons of the target nucleus in which it was produced, eventually leading to its suppression. This mechanism, firstly described in \cite{NAPLB207} has been experimentally observed, in particular at SPS \cite{NA38PLB466}\cite{NA50PLB553}\cite{NA50EPJC33}\cite{NA50EPJC48}.
\\
\begin{table}[tb]
    \centering
    \begin{tabular}{l|l|l|l}
        Experiment  &  Targets & $P_{beam}$ & $P_{targ}$ \\
                    &          &  (\gevc)   & (\gevc) \\
                    \hline
NA51 \cite{NA51PLB438}     & $p$, $d$ & 450 & 0 \\
NA50 \cite{NA50PLB553}     & Be,Al,Cu,Ag,W & 450 & 0 \\
NA50 \cite{NA50EPJC33}     & Be,Al,Cu,Ag,W & 450 & 0 \\
NA50 \cite{NA50EPJC48} & Be,Al,Cu,Ag,W,Pb & 400 & 0 \\
E288 \cite{E288PRL36}      & Be & 400 & 0 \\
E771 \cite{E771PLB374}     & Si & 800 & 0 \\
E789 \cite{E789PRD52}      & Au & 800 & 0 \\
PHENIX \cite{PHENIXPRC102} & Al, Au & 100 & 100 \\
PHENIX \cite{PHENIXPRL107} & Au & 100 & 100 \\
ALICE \cite{ALICEJHEP02}  & Pb  & 4000 & 1577 \\
LHCb \cite{LHCbPLB774}    & Pb  & 4000 & 1577 \\
    \end{tabular}
    \caption{Data used. All experiments operated $p$ induced reactions except \cite{PHENIXPRL107} which operated $d$-nucleus reactions. $P_{beam}$ and $P_{targ}$ correspond to the lab-frame beam and target momentum respectively.}
    \label{tab:ga}
\end{table}
The crossing time of a \ccbar pair in the rest frame of a nuclear target can be expressed as $t=L/v$, where $L$ corresponds to the length of nuclear matter traversed by the \ccbar pair and $v$ is the velocity of the \ccbar pair in the target rest frame, related to the \ccbar momentum by $p=\gamma mv$. Here, $m$ is the mass of the \ccbar system and $\gamma$ corresponds to its Lorentz factor. Thus, the proper time $\tau$ spent by the \ccbar pair in the target nucleus can be expressed as:
\begin{equation}
    \tau = \frac{t}{\gamma}=\frac{Lm}{p}=\frac{Lm}{\sqrt{p^2_z+p^2_T}}=\frac{Lm}{\sqrt{m^2_T\sinh^2y + p^2_T}}
    \label{eq:tau}
\end{equation}
where $y=0.5\times\ln((E+p_z)/(E-p_z))$ and $p_T=\sqrt{p^2-p^2_z}$ are the rapidity and the transverse momentum of the \ccbar state in the target frame respectively, and $m^2_T=m^2+p^2_T$ with $m$ the mass of the \ccbar state. For simplicity, we use, as a good approximation, $L = r (A^{1/3}-1)$, where $r=0.85$~fm and $A$ is the atomic mass number of the target. 
In the following, we study charmonium production as a function of $\tau$ for the datasets reported in Table \ref{tab:ga}, recorded with various targets at various energies. We assume that the small radius \ccbar pair, before it forms a charmonium, does not interact with the target nucleons on its path. Phenomenologically, in case of a sizeable charmonium formation time, and thanks to nuclear absorption, charmonium yields as a function of $\tau$ should exhibit a plateau, followed by a suppression. 
\begin{turnpage}
\begin{table*}[tb]
\begin{tabular}{lcccccccc}
 Experiment & $L$ & \sqsnn & $y_{CMS}$ & $y_{lab}$ & $y$   & $<p_T>$ & $x_2$ & $x_F$ \\ 
            & (fm)           & (\gev) &      &     &       & (\gevc) &    &   \\ 
\hline
NA51 \cite{NA51PLB438}     & $[0.00,0.22]$ & 29.1 & [$-0.4,0.6$] & [$3.03, 4.03$] & [$3.03,4.03$] & 1.0 & [$0.06,0.16$] & [$-0.09,0.15$] \\
NA50 \cite{NA50PLB553}     & $[0.92,3.98]$ & 29.1 & [$-0.4,0.6$] & [$3.03, 4.03$] & [$3.03,4.03$] & 1.0 & [$0.06,0.16$] & [$-0.09,0.15$] \\
NA50 \cite{NA50EPJC33}     & $[0.92,3.98]$ & 29.1 & [$-0.5,0.5$] & [$2.93, 3.93$] & [$2.93,3.93$] & 1.0 & [$0.07,0.18$] & [$-0.10,0.14$] \\
NA50 \cite{NA50EPJC48}     & $[0.92,4.18]$ &  27.4 & [$-0.425,0.575$] & [$2.95, 3.95$] & [$2.95,3.95$] & 1.0 & [$0.07,0.18$] & [$-0.10,0.14$] \\
E288 \cite{E288PRL36}      & $0.92$ & 27.4 & [$-0.28,0.32$] & [$3.09, 3.69$] & [$3.09,3.69$] & 1.0 & [$0.09,0.12$] & [$-0.07,0.08$] \\
E771 \cite{E771PLB374}     & $1.73$ & 38.8 & [$-0.55,1.32$] & [$3.16, 5.06$] & [$3.16,5.06$] & 1.0 & [$0.02,0.15$] & [$-0.10,0.30$] \\
E789 \cite{E789PRD52}      & $4.09$ & 38.8 & [$-0.17,0.78$] & [$3.54, 4.52$] & [$3.54,4.52$] & 1.0 & [$0.04,0.10$] & [$-0.03,0.15$] \\
PHENIX \cite{PHENIXPRC102} & $[1.70,4.09]$  & 200  & [$-2.2,-1.2$]  & [$-2.2,-1.2$] & [$3.16,4.16$] & 1.5 & [$0.06,0.12$] & [$-0.15,-0.05$] \\
PHENIX \cite{PHENIXPRL107} & $4.09$  & 200  & [$-2.2,-1.2$]  & [$-2.2, -1.2$] & [$3.16,4.16$] & 1.5 & [$0.06,0.12$] & [$-0.15,-0.05$] \\
ALICE \cite{ALICEJHEP02}  & $4.19$ &5020& [$-4.46,-2.96$] & [$-4.46, -2.96$] & [$4.13,5.63$] & 2.5 & [$0.01,0.06$] & [$-0.06,-0.01$] \\
LHCb \cite{LHCbPLB774}   & $4.19$ &5020& [$-4.00,-2.50$] & [$-4.00, -2.50$] & [$4.59,6.09$] & 2.5 & [$0.01,0.04$] & [$-0.04,-0.01$] \\
\end{tabular}
\caption{\label{tab:kinematics}Data used. $L=r(A^{1/3}-1)$, with $r=0.85$ fm, corresponds to the length of nuclear matter traversed by the \ccbar pair (numbers within brackets correspond to different targets), and \sqsnn to the center-of-mass energy of binary nucleon-nucleon collisions; $y_{CMS}$, $y_{lab}$ and $y$ are the rapidity of the \ccbar state in the center-of-mass, laboratory and target frames respectively, and $<P_T>$ its average transverse momentum as discussed in the text; $x_2$ and $x_F$ are the Bjorken-x and Feynman-x respectively as defined in eq.~\ref{eq:x}, considering smaller and larger $y$ values.}
\end{table*}
\end{turnpage}
Table \ref{tab:kinematics} provides kinematical information for the datasets used in this Letter. Since the \ccbar bound-state average transverse momentum $<p_T>$ is usually not reported, we follow the results given in \cite{PTJHEP10} and take:
\begin{itemize}
    \item $<p_T> \simeq 1.0\gevc$ for $\sqsnn<40\gev$,
    \item $<p_T> \simeq 1.5\gevc$ for $\sqsnn=200\gev$,
    \item $<p_T> \simeq 2.5\gevc$ for $\sqsnn>2 \tev$.
\end{itemize}
Bjorken-x $x_2$ and Feyman-x $x_F$ are calculated following eq.~\ref{eq:x}, taking $m=3.097\gevcc$ \cite{PDG:2020yd}, the mass of the \jpsi:
\begin{equation}\label{eq:x}
    x_2 = \frac{m_T}{\sqsnn}e^{-y_{CMS}},~~~
    x_F = \frac{2 m_T}{\sqsnn}\sinh(y_{CMS})
\end{equation}
Because charmonia may suffer several cold nuclear matter effects, the data used in this letter are chosen to cover kinematical regions where  nuclear effects but nuclear absorption do not apply, or, at least, are expected to be small. 
The criteria are:
\begin{itemize}
    \item $x_F$ must be close to zero, far from the energy loss \cite{ELJHEP03} and saturation regimes,
    \item $x_2$ must belong to the region $[10^{-2},10^{-1}]$, close to the transition between nPDF shadowing and anti-shadowing regions \cite{PDFEPJC77,PDFPRD93}, where those effects are expected to be small,
    \item quarkonium interaction with comoving hadrons must be small, limiting the use of \psip to the low energy (fixed-target) data samples \cite{Ferreiro:2014bia}. 
\end{itemize}
Relevant data are also required to extend over reduced $\tau$ ranges corresponding to reduced rapidity ranges. In the following, data points are reported with uncertainties on the $\tau$ values corresponding to the rapidity ranges in which data were recorded. 
\\
We propose the following parametrization of the nuclear absorption \cite{NPB345,PRC87} based on the proper time spent in the nucleus by the quarkonia (or \ccbar precursor):
\begin{equation}\label{eq:abs}
    \sigma^\ccbar_{pA} = A~\sigma^\ccbar_{pp}~\exp(-\rho_0\sigma_{abs}~\beta\gamma c\tau)
\end{equation}
where $\sigma^\ccbar_{pA}$ is the charmonium production cross section in $p$A collisions, $A$ the atomic mass number of the target nucleus,  $\sigma^\ccbar_{pp}$ the charmonium production cross section in $pp$ collisions, $\sigma_{abs}$ the charmonium absorption (or break-up) cross section, $\rho_0=0.170$ fm$^{-3}$ the nuclear density and  $\beta\gamma c\tau$ (with $\beta=v/c$ and $c$ the speed of light) is the length of nuclear matter traversed by the \ccbar pair. Assuming that, before charmonium state formation time $\tau_0$, the small radius \ccbar pair does not interact with the target nucleons on its path, we propose, first, based on eq.~\ref{eq:abs}, to introduce $\tau_0$ in the step function:
\begin{equation}\label{eq:abs2}
    \frac{\sigma^\ccbar_{pA}}{A\sigma^\ccbar_{pp}} = \left\{
    \begin{array}{lc}
         1 &  \hbox{if } \tau<\tau_0\\
         \exp(-\rho_0\sigma_{abs}~\beta\gamma c(\tau-\tau_0))&\hbox{if } \tau>\tau_0 
    \end{array}
    \right .
\end{equation}
where all \ccbar bound states start suffering nuclear absorption when reaching $\tau_0$.
\\
\begin{figure}[t]
\includegraphics[width=0.75\textwidth]{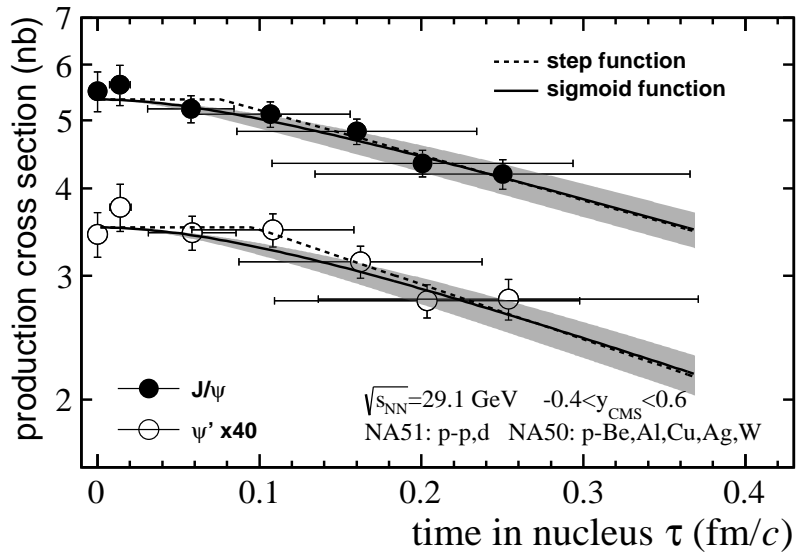}
\caption{\label{fig:CrossSections} \jpsi and \psip $A$-normalised production cross section ($\sigma^\ccbar_{pA}/A$) from \cite{NA50PLB553,NA51PLB438} as functions of $\tau$, defined in eq.~\ref{eq:tau}. The \psip cross sections have been scaled by a factor 40 to ease comparison. Dashed lines correspond to the step function (eq.~\ref{eq:abs2}), plain lines to the sigmoid function (eq.~\ref{eq:exact}), all parameters being determined with fits based on eq.~\ref{eq:abs2}. Grey bands show the effects of $\tau_0^\jpsi$ and $\tau_0^\psip$ uncertainties on the sigmoid function.}
\end{figure}
Figure \ref{fig:CrossSections} shows the \jpsi and \psip production cross sections measured by the NA51 \cite{NA51PLB438} and NA50 \cite{NA50PLB553} experiments at $\sqsnn=29.1 \gev$, as functions of $\tau$ as defined in eq.~\ref{eq:tau}. In both cases, a structure appears, made of a plateau, followed by a suppression. 
Taking $\sigma^\ccbar_{pp}$, $\sigma_{abs}$ and $\tau_0$ as free parameters, $\chi^2$-minimization fits based on eq.~\ref{eq:abs2} give $\tau_0^\jpsi=0.08\pm0.04$~fm/$c$ and $\tau^\psip_0=0.10\pm0.04$~fm/$c$, for the \jpsi and the \psip respectively. For completeness, the values of $\sigma^\ccbar_{pp}$ and $\sigma_{abs}$ are reported in Table \ref{tab:res}.
\\
More realistically, considering that \ccbar hadronization follows the standard decay law $dN^\ccbar=-\lambda N^\ccbar dt$, with $\lambda=1/\gamma\tau_0$ and $t=\gamma\tau$, the charmonium survival probability follows the sigmoid function:
\begin{eqnarray}\label{eq:exact}
    \frac{\sigma^\ccbar_{pA}}{A\sigma^\ccbar_{pp}} & = & \int_0^{L/\beta c} \lambda e^{-\lambda t}e^{-\rho_0\sigma(L-\beta ct)}dt + \int_{L/\beta c}^\infty \lambda e^{-\lambda t}dt\nonumber \\
    & = & 
    \frac{e^{-\rho_0\sigma_{abs}\beta\gamma c\tau}}{1-\rho_0\sigma_{abs}\beta\gamma c\tau_0}\nonumber \\
    & & - e^{-\tau/\tau_0}\left ( \frac{1}{1-\rho_0\sigma_{abs}\beta\gamma c\tau_0} - 1\right )
\end{eqnarray}
tending to the step function (eq.~\ref{eq:abs2}) for small and large values of $\tau$ as illustrated on Figure~\ref{fig:CrossSections}. Note that, due to the interplay of the decay and absorption terms in eq.~\ref{eq:exact}, current experimental results do not permit to obtain reliable minimizations without constraining the parameters. We therefore take the results obtained with the step function (eq.~\ref{eq:abs2}) as fixed input parameters for the sigmoid function (eq.~\ref{eq:exact}). 
\\
We now consider experimental results recorded in different experimental conditions, as reported in Table \ref{tab:ga}. Because quarkonium cross section depends on center-of-mass energy, cross section ratios, such as the nuclear modification factor $R_{AB}$ and the \psip over \jpsi cross sections ratio $\sigma_{\psip}/\sigma_{\jpsi}$ are appropriate quantities to compare data from various experiments at various energies. 
\begin{figure}[t]
\includegraphics[width=0.75\textwidth]{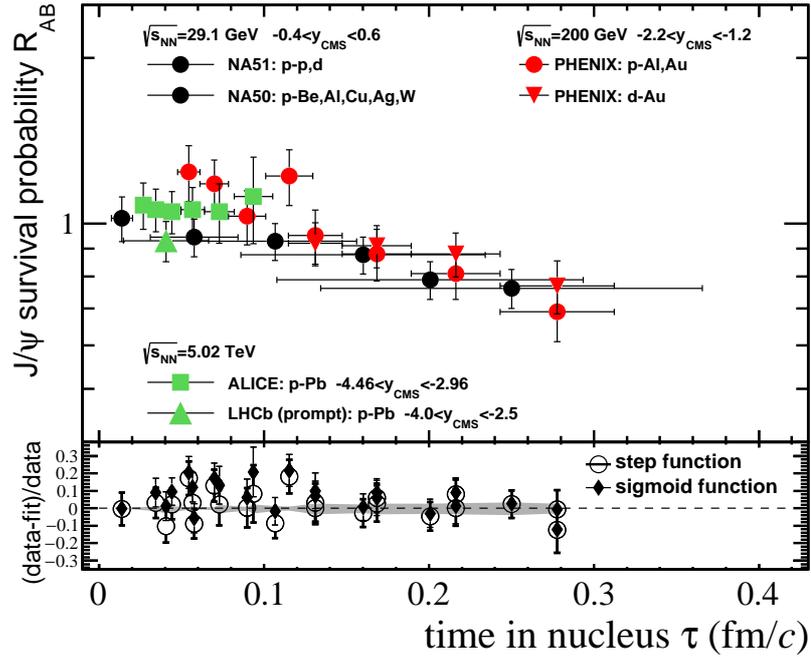}
\caption{\label{fig:NMF} Top: \jpsi nuclear modification factor $R_{AB}$ from \cite{NA51PLB438,NA50PLB553,PHENIXPRC102,ALICEJHEP02,LHCbPLB774} as a function of $\tau$ as defined in eq.~\ref{eq:tau}, with $A=1$ (proton beam) or $A=2$ (deuteron beam); Bottom: Relative difference between data and eq.~\ref{eq:abs2} step function (open circles), between data and eq.~\ref{eq:exact} sigmoid function (close diamonds), all parameters being determined with fits based on eq.~\ref{eq:abs2}.The Grey band shows the effects of $\tau_0^\jpsi$ uncertainties on the sigmoid function.}
\end{figure}
Figure~\ref{fig:NMF} shows the \jpsi nuclear modification factor $R_{AB}$ as a function of $\tau$ for the SPS, RHIC and LHC experimental data, listed in Tables \ref{tab:ga} and \ref{tab:kinematics}, from $\sqsnn=29.1\gev$ to $\sqsnn=5.02\tev$ proton-nucleus collisions. In our framework, $R_{AB}$ is a proxy for the probability of a \jpsi to survive when traversing the nucleus. At LHC energies, the ALICE experiment reports on the inclusive \jpsi cross section (including an additional yield of $10 - 15\%$ due to \jpsi-from-$B$ hadron decay contribution), while the LHCb experiment reports on the prompt \jpsi cross section (excluding \jpsi-from-$B$ hadron decays). At lower energies, where contributions from $B$ hadron decays are small, all experiments report on the inclusive \jpsi cross sections. The nuclear modification factors for the NA50 and NA51 experiments have been calculated with:
\begin{equation}
    R_{AB}=\frac{\sigma^\jpsi_{AB}}{AB~\sigma^\jpsi_{pp}}
\end{equation}
where the uncertainties on the NA51 $pp$ \jpsi cross section have been propagated to the ratio. As in Figure~\ref{fig:CrossSections}, a plateau is observed for small values of $\tau$. Moreover, although recorded at very different energies, NA50 and PHENIX data, in the region $\tau>0.1$~fm/$c$, follow a similar trend, consistent with a suppression scenario depending on geometrical effect such as nuclear absorption.
The results of a fit based on eq.~\ref{eq:abs2} are reported in Table \ref{tab:res} with  $\tau_0^\jpsi=0.10\pm 0.02$~fm/$c$, in agreement with the value obtained for Figure \ref{fig:CrossSections}. Beside, since experimental data have been recorded in different kinematical regimes, the $\beta\gamma$ factor depends on the data sample, preventing reporting the fit results on the plot. We instead report on the $(data-fit)/data$ ratio where, as for Figure \ref{fig:CrossSections}, the results obtained with eq.~\ref{eq:abs2} are used as fixed input parameters for the sigmoid function. The corresponding $\chi$-square per degree of freedom, $\chi^2_4/ndf=0.65$ and $\chi^2_5/ndf=0.90$ for the step (eq.~\ref{eq:abs2}) and sigmoid (eq.~\ref{eq:exact}) functions respectively, indicate good agreement between data and fit.  
\\
Figure \ref{fig:Ratios} shows the \psip/\jpsi cross section ratio as a function of $\tau$ for several data collected at various energies by the CERN NA51 \cite{NA51PLB438} and NA50 \cite{NA50PLB553} experiments, and the Fermilab E288 \cite{E288PRL36}, E771 \cite{E771PLB374} and E789 \cite{E789PRD52} experiments.  PHENIX and LHC data are not considered here since, because of the large center-of-mass energy, \psip production is expected to be significantly suppressed by interacting with comovers. Here again, a plateau is observed for small values of $\tau$. A fit based on eq.~\ref{eq:abs2}, with the absorption cross section $\Delta\sigma_{abs}=\sigma_{abs}^{\psip}-\sigma_{abs}^\jpsi$ and assuming $\tau_0^\jpsi=\tau^\psip_0=\tau_0$, gives $\tau_0=0.13\pm 0.3$~fm/$c$, in good agreement with the results obtained from Figure \ref{fig:CrossSections} and  \ref{fig:NMF}. The $(data-fit)/data$ ratio, as for Figure~\ref{fig:NMF}, is obtained with both the step (eq.~\ref{eq:abs2}) and sigmoid (eq.~\ref{eq:exact}) functions. The corresponding $\chi$-squares per degree of freedom, $\chi^2_4/ndf=0.86$ and $\chi^2_5/ndf=1.37$ for the step and sigmoid functions respectively, also indicate good agreement between data and fit.
\begin{table}[h]
    \centering
    \begin{tabular}{ccccc}
        Fig.  & Quantity & Fit with eq.~\ref{eq:abs2} 
                & $\frac{\chi^2_{4}}{ndf}$ & $\frac{\chi^2_{5}}{ndf}$\\
        \hline
          \multirow{3}{*}{1} & \multirow{3}{*}{$\sigma^\jpsi_{pA}/A$}    
                             & $\sigma^\jpsi_{pp}=5.35 \pm 0.17$ nb 
                             & \multirow{3}{*}{0.45} & \multirow{3}{*}{0.26}\\
              & & $\sigma^\jpsi_{abs}=4.3\pm 1.2$ mb &&\\ 
              & &  $\tau_0^\jpsi=0.08\pm0.04$~fm/$c$&&\\
                \hline
          \multirow{4}{*}{1} & \multirow{4}{*}{$\sigma^\psip_{pA}/A$}    
              & $\sigma^\psip_{pp}=88.9 \pm 3.4$ pb 
              & \multirow{3}{*}{0.57} & \multirow{3}{*}{0.52} \\
              & & $\sigma^\psip_{abs}=5.3\pm 1.7$ mb &&\\
              & & $\tau^\psip_0=0.10\pm0.04$~fm/$c$ &&\\
                \hline
          \multirow{4}{*}{2}   & \multirow{3}{*}{$\frac{\sigma^\jpsi_{AB}}{AB~\sigma^\jpsi_{pp}}$} 
          & $norm.=1.03\pm 0.03$ & \multirow{3}{*}{0.65} 
          &\multirow{3}{*}{0.90}\\
              &  & $\sigma^\jpsi_{abs}=6.1\pm 1.4$ mb &&\\
              &  & $\tau_0^\jpsi=0.10\pm 0.02$~fm/$c$ &&\\
                \hline
          \multirow{4}{*}{3}   &  
          \multirow{3}{*}{$\frac{Br_{\mu\mu}^\psip\sigma^\psip}{Br_{\mu\mu}^\jpsi\sigma^\jpsi}$} & $norm.(\%)=1.7\pm0.2$ & \multirow{3}{*}{0.86} & \multirow{3}{*}{1.37}\\
           && $\Delta\sigma_{abs}=2.7\pm 0.7$ mb &&\\
              &  & $\tau_0=0.13\pm 0.03$~fm/$c$ &&\\
    \end{tabular}
    \caption{Results of the fits obtained with eq.~\ref{eq:abs2}; $\chi^2_4/ndf$ corresponds to the $\chi$-squared per degree of freedom calculated with the step function (eq.~\ref{eq:abs2}), $\chi^2_5/ndf$ to the same quantity calculated with the sigmoid function (eq.~\ref{eq:exact}).}
    \label{tab:res}
\end{table}
In conclusion, when studying \ccbar bound state productions as a function of $\tau$, the proper time spent by the \ccbar pair in nuclear matter, for data recorded with various targets at different energies, a structure appears, made of a plateau up to the time $\tau_0\sim0.1$ fm/$c$, followed by a suppression. Although current experimental uncertainties prevent drawing any firm conclusion, this suppression pattern, if confirmed, could provide important information on the \ccbar pair hadronization into charmonium bound state, opening the gate to other measurements of this kind. In order to precisely test this scenario, an experimental program collecting large statistical samples with various targets in the appropriate kinematical region would certainly offer a privileged configuration. 
%\begin{acknowledgment}
%We wish to thank Benjamin Audurier for very fruitful discussions on quarkonium production mechanisms, and acknowledge the contribution of Raïssa Costa Barroso in studying the sensitivity of this approach with $y$ and $p_T$.   
%\end{acknowledgment}
% References
\addcontentsline{toc}{section}{References}
\providecommand{\noopsort}[1]{}\providecommand{\singleletter}[1]{#1}%

\begin{thebibliography}{28}%
\makeatletter
\providecommand \@ifxundefined [1]{%
 \@ifx{#1\undefined}
}%
\providecommand \@ifnum [1]{%
 \ifnum #1\expandafter \@firstoftwo
 \else \expandafter \@secondoftwo
 \fi
}%
\providecommand \@ifx [1]{%
 \ifx #1\expandafter \@firstoftwo
 \else \expandafter \@secondoftwo
 \fi
}%
\providecommand \natexlab [1]{#1}%
\providecommand \enquote  [1]{``#1''}%
\providecommand \bibnamefont  [1]{#1}%
\providecommand \bibfnamefont [1]{#1}%
\providecommand \citenamefont [1]{#1}%
\providecommand \href@noop [0]{\@secondoftwo}%
\providecommand \href [0]{\begingroup \@sanitize@url \@href}%
\providecommand \@href[1]{\@@startlink{#1}\@@href}%
\providecommand \@@href[1]{\endgroup#1\@@endlink}%
\providecommand \@sanitize@url [0]{\catcode `\\12\catcode `\$12\catcode
  `\&12\catcode `\#12\catcode `\^12\catcode `\_12\catcode `\%12\relax}%
\providecommand \@@startlink[1]{}%
\providecommand \@@endlink[0]{}%
\providecommand \url  [0]{\begingroup\@sanitize@url \@url }%
\providecommand \@url [1]{\endgroup\@href {#1}{\urlprefix }}%
\providecommand \urlprefix  [0]{URL }%
\providecommand \Eprint [0]{\href }%
\providecommand \doibase [0]{https://doi.org/}%
\providecommand \selectlanguage [0]{\@gobble}%
\providecommand \bibinfo  [0]{\@secondoftwo}%
\providecommand \bibfield  [0]{\@secondoftwo}%
\providecommand \translation [1]{[#1]}%
\providecommand \BibitemOpen [0]{}%
\providecommand \bibitemStop [0]{}%
\providecommand \bibitemNoStop [0]{.\EOS\space}%
\providecommand \EOS [0]{\spacefactor3000\relax}%
\providecommand \BibitemShut  [1]{\csname bibitem#1\endcsname}%
\let\auto@bib@innerbib\@empty
%</preamble>
\bibitem {JPLPR889}%
  \BibitemOpen
  \bibfield  {author} {\bibinfo {author} {\bibfnamefont {J.~P.}\ \bibnamefont
  {Lansberg}},\ }\href {https://doi.org/10.1016/j.physrep.2020.08.007}
  {\bibfield  {journal} {\bibinfo  {journal} {Phys. Rep.}\ }\textbf {\bibinfo
  {volume} {889}},\ \bibinfo {pages} {1} (\bibinfo {year} {2020})}\BibitemShut
  {NoStop}%
\bibitem {NAPLB207}%
  \BibitemOpen
  \bibfield  {author} {\bibinfo {author} {\bibfnamefont {C.}~\bibnamefont
  {Gerschel}}\ and\ \bibinfo {author} {\bibfnamefont {J.}~\bibnamefont
  {Hüfner}},\ }\href {https://doi.org/10.1016/0370-2693(88)90570-9} {\bibfield
   {journal} {\bibinfo  {journal} {Phys. Lett. B}\ }\textbf {\bibinfo {volume}
  {207}},\ \bibinfo {pages} {253} (\bibinfo {year} {1988})}\BibitemShut
  {NoStop}%
\bibitem {ELJHEP03}%
  \BibitemOpen
  \bibfield  {author} {\bibinfo {author} {\bibfnamefont {F.}~\bibnamefont
  {Arleo}}\ and\ \bibinfo {author} {\bibfnamefont {S.}~\bibnamefont
  {Peign\'e}},\ }\href {https://doi.org/10.1007/JHEP03(2013)122} {\bibfield
  {journal} {\bibinfo  {journal} {J. High Energ. Phys.}\ }\textbf {\bibinfo
  {volume} {03}},\ \bibinfo {pages} {122} (\bibinfo {year} {2013})}\BibitemShut {NoStop}%
\bibitem {CapellaPLB393}%
  \BibitemOpen
  \bibfield  {author} {\bibinfo {author} {\bibfnamefont {A.}~\bibnamefont
  {Capella}} \emph {et~al.},\ }\href
  {https://doi.org/10.1016/S0370-2693(96)01650-4} {\bibfield  {journal}
  {\bibinfo  {journal} {Phys. Lett. B}\ }\textbf {\bibinfo {volume} {393}},\
  \bibinfo {pages} {431} (\bibinfo {year} {1997})}\BibitemShut {NoStop}%
\bibitem {GavinPRL78}%
  \BibitemOpen
  \bibfield  {author} {\bibinfo {author} {\bibfnamefont {S.}~\bibnamefont
  {Gavin}}\ and\ \bibinfo {author} {\bibfnamefont {R.}~\bibnamefont {Vogt}},\
  }\href {https://doi.org/10.1103/PhysRevLett.78.1006} {\bibfield  {journal}
  {\bibinfo  {journal} {Phys. Rev. Lett.}\ }\textbf {\bibinfo {volume} {78}},\
  \bibinfo {pages} {1006} (\bibinfo {year} {1997})}\BibitemShut {NoStop}%
\bibitem {ArmestoPLB430}%
  \BibitemOpen
  \bibfield  {author} {\bibinfo {author} {\bibfnamefont {N.}~\bibnamefont
  {Armesto}}\ and\ \bibinfo {author} {\bibfnamefont {A.}~\bibnamefont
  {Capella}},\ }\href {https://doi.org/10.1016/S0370-2693(98)00487-0}
  {\bibfield  {journal} {\bibinfo  {journal} {Phys. Lett. B}\ }\textbf
  {\bibinfo {volume} {430}},\ \bibinfo {pages} {23} (\bibinfo {year}
  {1998})}\BibitemShut {NoStop}%
\bibitem {CapellaPRL85}%
  \BibitemOpen
  \bibfield  {author} {\bibinfo {author} {\bibfnamefont {A.}~\bibnamefont
  {Capella}} \emph {et~al.},\ }\href
  {https://doi.org/10.1103/PhysRevLett.85.2080} {\bibfield  {journal} {\bibinfo
   {journal} {Phys. Rev. Lett.}\ }\textbf {\bibinfo {volume} {85}},\ \bibinfo
  {pages} {2080} (\bibinfo {year} {2000})}\BibitemShut {NoStop}%
\bibitem {PDFEPJC77}%
  \BibitemOpen
  \bibfield  {author} {\bibinfo {author} {\bibfnamefont {K.~J.}\ \bibnamefont
  {Eskola}} \emph {et~al.},\ }\href
  {https://doi.org/10.1140/epjc/s10052-017-4725-9} {\bibfield  {journal}
  {\bibinfo  {journal} {Eur. Phys. J. C}\ }\textbf {\bibinfo {volume} {77}},\
  \bibinfo {pages} {163} (\bibinfo {year} {2017})}\BibitemShut {NoStop}%
\bibitem {PDFPRD93}%
  \BibitemOpen
  \bibfield  {author} {\bibinfo {author} {\bibfnamefont {K.}~\bibnamefont
  {Kovarik}} \emph {et~al.},\ }\href
  {https://doi.org/10.1103/PhysRevD.93.085037} {\bibfield  {journal} {\bibinfo
  {journal} {Phys. Rev. D}\ }\textbf {\bibinfo {volume} {93}},\ \bibinfo
  {pages} {085037} (\bibinfo {year} {2016})}\BibitemShut {NoStop}%
\bibitem {NPB345}%
  \BibitemOpen
  \bibfield  {author} {\bibinfo {author} {\bibfnamefont {S.}~\bibnamefont
  {Gavin}}\ and\ \bibinfo {author} {\bibfnamefont {R.}~\bibnamefont {Vogt}},\
  }\href {https://doi.org/10.1016/0550-3213(90)90610-P} {\bibfield  {journal}
  {\bibinfo  {journal} {Nucl. Phys. B}\ }\textbf {\bibinfo {volume} {345}},\
  \bibinfo {pages} {104} (\bibinfo {year} {1990})}\BibitemShut {NoStop}%
\bibitem {PRC87}%
  \BibitemOpen
  \bibfield  {author} {\bibinfo {author} {\bibfnamefont {D.~C.}\ \bibnamefont
  {McGlinchey}} \emph {et~al.},\ }\href
  {https://doi.org/10.1103/PhysRevC.87.054910} {\bibfield  {journal} {\bibinfo
  {journal} {Phys. Rev. C}\ }\textbf {\bibinfo {volume} {87}},\ \bibinfo
  {pages} {054910} (\bibinfo {year} {2013})}\BibitemShut {NoStop}%
\bibitem {LOUR09}%
  \BibitemOpen
  \bibfield  {author} {\bibinfo {author} {\bibfnamefont {C.}\ \bibnamefont
  {Lourenco}} \emph {et~al.},\ }\href
  {https://doi.org/10.1088/1126-6708/2009/02/014} {\bibfield  {journal} {\bibinfo
  {journal} {J. High Energ. Phys.}\ }\textbf {\bibinfo {volume} {02}},\ \bibinfo
  {pages} {014} (\bibinfo {year} {2009})}\BibitemShut {NoStop}%
\bibitem {KOPPRD44}%
  \BibitemOpen
  \bibfield  {author} {\bibinfo {author} {\bibfnamefont {B.~Z.}\ \bibnamefont
  {Kopeliovich}}\ and\ \bibinfo {author} {\bibfnamefont {B.~G.}\ \bibnamefont
  {Zakharov}},\ }\href {https://doi.org/10.1103/PhysRevD.44.3466} {\bibfield
  {journal} {\bibinfo  {journal} {Phys. Rev. D}\ }\textbf {\bibinfo {volume}
  {44}},\ \bibinfo {pages} {3466} (\bibinfo {year} {1991})}\BibitemShut
  {NoStop}%
\bibitem {BRODPLB206}%
  \BibitemOpen
  \bibfield  {author} {\bibinfo {author} {\bibfnamefont {S.~J.}\ \bibnamefont
  {Brodsky}}\ and\ \bibinfo {author} {\bibfnamefont {A.~H.}\ \bibnamefont
  {Mueller}},\ }\href {https://doi.org/10.1016/0370-2693(88)90719-8} {\bibfield
   {journal} {\bibinfo  {journal} {Phys. Lett. B}\ }\textbf {\bibinfo {volume}
  {206}},\ \bibinfo {pages} {685} (\bibinfo {year} {1988})}\BibitemShut
  {NoStop}%
\bibitem {NA38PLB466}%
  \BibitemOpen
  \bibfield  {author} {\bibinfo {author} {\bibfnamefont {M.~C.}\ \bibnamefont
  {Abreu}} \emph {et~al.} (\bibinfo {collaboration} {NA38}),\ }\href
  {https://doi.org/10.1016/S0370-2693(99)00057-X} {\bibfield  {journal}
  {\bibinfo  {journal} {Phys. Lett. B}\ }\textbf {\bibinfo {volume} {449}},\
  \bibinfo {pages} {128} (\bibinfo {year} {1999})}\BibitemShut {NoStop}%
\bibitem {NA50PLB553}%
  \BibitemOpen
  \bibfield  {author} {\bibinfo {author} {\bibfnamefont {B.}~\bibnamefont
  {Alessandro}} \emph {et~al.} (\bibinfo {collaboration} {NA50}),\ }\href
  {https://doi.org/10.1016/S0370-2693(02)03265-3} {\bibfield  {journal}
  {\bibinfo  {journal} {Phys. Lett. B}\ }\textbf {\bibinfo {volume} {553}},\
  \bibinfo {pages} {167} (\bibinfo {year} {2003})}\BibitemShut {NoStop}%
\bibitem {NA50EPJC33}%
  \BibitemOpen
  \bibfield  {author} {\bibinfo {author} {\bibfnamefont {B.}~\bibnamefont
  {Alessandro}} \emph {et~al.} (\bibinfo {collaboration} {NA50}),\ }\href
  {https://doi.org/10.1140/epjc/s2003-01539-y} {\bibfield  {journal} {\bibinfo
  {journal} {Eur. Phys. J. C}\ }\textbf {\bibinfo {volume} {33}},\ \bibinfo
  {pages} {31} (\bibinfo {year} {2004})}\BibitemShut {NoStop}%
\bibitem {NA50EPJC48}%
  \BibitemOpen
  \bibfield  {author} {\bibinfo {author} {\bibfnamefont {B.}~\bibnamefont
  {Alessandro}} \emph {et~al.} (\bibinfo {collaboration} {NA50}),\ }\href
  {https://doi.org/10.1140/epjc/s10052-006-0079-4} {\bibfield  {journal}
  {\bibinfo  {journal} {Eur. Phys. J. C}\ }\textbf {\bibinfo {volume} {48}},\
  \bibinfo {pages} {329} (\bibinfo {year} {2006})}\BibitemShut {NoStop}%
\bibitem {NA51PLB438}%
  \BibitemOpen
  \bibfield  {author} {\bibinfo {author} {\bibfnamefont {M.~C.}\ \bibnamefont
  {Abreu}} \emph {et~al.} (\bibinfo {collaboration} {NA51}),\ }\href
  {https://doi.org/https://doi.org/10.1016/S0370-2693(98)01014-4} {\bibfield
  {journal} {\bibinfo  {journal} {Phys. Lett. B}\ }\textbf {\bibinfo {volume}
  {438}},\ \bibinfo {pages} {35} (\bibinfo {year} {1998})}\BibitemShut
  {NoStop}%
\bibitem {E288PRL36}%
  \BibitemOpen
  \bibfield  {author} {\bibinfo {author} {\bibfnamefont {H.~D.}\ \bibnamefont
  {Snyder}} \emph {et~al.},\ }\href
  {https://doi.org/10.1103/PhysRevLett.36.1415} {\bibfield  {journal} {\bibinfo
   {journal} {Phys. Rev. Lett.}\ }\textbf {\bibinfo {volume} {36}},\ \bibinfo
  {pages} {1415} (\bibinfo {year} {1976})}\BibitemShut {NoStop}%
\bibitem {E771PLB374}%
  \BibitemOpen
  \bibfield  {author} {\bibinfo {author} {\bibfnamefont {T.}~\bibnamefont
  {Alexopoulos}} \emph {et~al.} (\bibinfo {collaboration} {E771}),\ }\href
  {https://doi.org/10.1016/0370-2693(96)00256-0} {\bibfield  {journal}
  {\bibinfo  {journal} {Phys. Lett. B}\ }\textbf {\bibinfo {volume} {374}},\
  \bibinfo {pages} {271} (\bibinfo {year} {1996})}\BibitemShut {NoStop}%
\bibitem {E789PRD52}%
  \BibitemOpen
  \bibfield  {author} {\bibinfo {author} {\bibfnamefont {M.~H.}\ \bibnamefont
  {Schub}} \emph {et~al.} (\bibinfo {collaboration} {E789}),\ }\href
  {https://doi.org/10.1103/PhysRevD.52.1307} {\bibfield  {journal} {\bibinfo
  {journal} {Phys. Rev. D}\ }\textbf {\bibinfo {volume} {52}},\ \bibinfo
  {pages} {1307} (\bibinfo {year} {1995})}\BibitemShut {NoStop}%
\bibitem {PHENIXPRC102}%
  \BibitemOpen
  \bibfield  {author} {\bibinfo {author} {\bibfnamefont {U.}~\bibnamefont
  {Acharya}} \emph {et~al.} (\bibinfo {collaboration} {PHENIX}),\ }\href
  {https://doi.org/10.1103/PhysRevC.102.014902} {\bibfield  {journal} {\bibinfo
   {journal} {Phys. Rev. C}\ }\textbf {\bibinfo {volume} {102}},\ \bibinfo
  {pages} {014902} (\bibinfo {year} {2020})}\BibitemShut {NoStop}%
\bibitem {PHENIXPRL107}%
  \BibitemOpen
  \bibfield  {author} {\bibinfo {author} {\bibfnamefont {A.}~\bibnamefont
  {Adare}} \emph {et~al.} (\bibinfo {collaboration} {PHENIX}),\ }\href
  {https://doi.org/10.1103/PhysRevLett.107.142301} {\bibfield  {journal}
  {\bibinfo  {journal} {Phys. Rev. Lett.}\ }\textbf {\bibinfo {volume} {107}},\
  \bibinfo {pages} {142301} (\bibinfo {year} {2011})}\BibitemShut {NoStop}%
\bibitem {ALICEJHEP02}%
  \BibitemOpen
  \bibfield  {author} {\bibinfo {author} {\bibfnamefont {B.}~\bibnamefont
  {Abelev}} \emph {et~al.} (\bibinfo {collaboration} {ALICE}),\ }\href
  {https://doi.org/10.1007/JHEP12(2014)073} {\bibfield  {journal} {\bibinfo
  {journal} {J. High Energ. Phys.}\ }\textbf {\bibinfo {volume} {2014}},\
  \bibinfo {pages} {73} (\bibinfo {year} {2014})}\BibitemShut {NoStop}%
\bibitem {LHCbPLB774}%
  \BibitemOpen
  \bibfield  {author} {\bibinfo {author} {\bibfnamefont {R.}~\bibnamefont
  {Aaij}} \emph {et~al.} (\bibinfo {collaboration} {LHCb}),\ }\href
  {https://doi.org/10.1016/j.physletb.2017.09.058} {\bibfield  {journal}
  {\bibinfo  {journal} {Phys. Lett. B}\ }\textbf {\bibinfo {volume} {774}},\
  \bibinfo {pages} {159 } (\bibinfo {year} {2017})}\BibitemShut {NoStop}%
\bibitem {PTJHEP10}%
  \BibitemOpen
  \bibfield  {author} {\bibinfo {author} {\bibfnamefont {S.}~\bibnamefont
  {Acharya}} \emph {et~al.} (\bibinfo {collaboration} {ALICE}),\ }\href
  {https://doi.org/10.1007/JHEP10(2019)084} {\bibfield  {journal} {\bibinfo
  {journal} {J. High Energ. Phys.}\ }\textbf {\bibinfo {volume} {2019}},\
  \bibinfo {pages} {84} (\bibinfo {year} {2019})}\BibitemShut {NoStop}%
\bibitem {PDG:2020yd}%
  \BibitemOpen
  \bibfield  {author} {\bibinfo {author} {\bibfnamefont {P.}~\bibnamefont
  {Zyla}} \emph {et~al.} (\bibinfo {collaboration} {Particle Data Group}),\
  }\href {https://doi.org/10.1093/ptep/ptaa104} {\bibfield  {journal} {\bibinfo
   {journal} {Prog. Theor. Exp. Phys.}\ }\textbf {\bibinfo {volume} {2020}},\
  \bibinfo {pages} {083C01} (\bibinfo {year} {2020})}\BibitemShut {NoStop}%
\bibitem {Ferreiro:2014bia}%
  \BibitemOpen
  \bibfield  {author} {\bibinfo {author} {\bibfnamefont {E.~G.}\ \bibnamefont
  {Ferreiro}},\ }\href {https://doi.org/10.1016/j.physletb.2014.02.011}
  {\bibfield  {journal} {\bibinfo  {journal} {Phys. Lett. B}\ }\textbf
  {\bibinfo {volume} {731}},\ \bibinfo {pages} {57} (\bibinfo {year}
  {2014})}\BibitemShut {NoStop}%
\end{thebibliography}
\end{document}